\documentclass[11pt,a4paper]{article}
\pdfoutput=1
\usepackage{jheppub}
\usepackage{graphics}
\usepackage{stmaryrd}
\usepackage{amsfonts}
\usepackage{mathrsfs}
\usepackage{color}
\usepackage{bm}
\usepackage{amsmath}
\usepackage{chemarrow}

\newcommand{\fig}[1]{Fig.\ref{#1}}
\def\be{\begin{equation}}
\def\ee{\end{aligned}\ea}
\def\ba{\begin{eqnarray}}
\def\ea{\end{eqnarray}}

\def\nn{\nonumber}

\newcommand{\eq}[1]{(\ref{#1})}

\def\q{\theta}      \def\p {\pi} \def\a {\alpha} \def\s {\sigma} \def\d {\delta} \def\f {\phi} \def\g {\gamma}  \def\j {\varphi} \def\k {\kappa} \def\l {\lambda} \def\z {\zeta} \def\x {\xi}  \def\b {\beta}   \def\pd {\partial}\def\p {\pi} \def \inf {\infty}  \def \e { \varepsilon}
 \def\W{\Omega}     \def\S {\Sigma}  \def\F {\Phi} \def\G {\Gamma}     \def\.{\cdot}
\def\math {\mathcal}
\title{Gedanken experiments at high-order approximation: Kerr black hole cannot be overspun}
\author{Aofei Sang$^{ab}$}
\author{and Jie Jiang$^{ab}$\footnote{Corresponding author.}}

\affiliation[a]{College of Education for the Future, Beijing Normal University, Zhuhai 519087, China}

\affiliation[b]{Department of Physics, Beijing Normal University, Beijing, 100875, China}

\emailAdd{202021140021@mail.bnu.edu.cn, jiejiang@mail.bnu.edu.cn}

\abstract{Sorce and Wald proposed a new version of gedanken experiments to examine the weak cosmic censorship conjecture (WCCC) in Kerr-Newmann black holes. However, their discussion only includes the second-order approximation of perturbation and there exists an optimal condition such that the validity of the WCCC is determined by the higher-order approximations. Therefore, in this paper, we extended their discussions into the high-order approximations to study the WCCC in a nearly extremal Kerr black hole. After assuming that the spacetime satisfies the stability condition and the perturbation matter fields satisfy the null energy condition, based on the Noether charge method by Iyer and Wald, we completely calculate the first four order perturbation inequalities and discuss the corresponding gedanken experiment to overspin the Kerr black hole. As a result, we find that the nearly extremal Kerr black holes cannot be destroyed under the fourth-order approximation of perturbation. Then, by using the mathematical induction, we strictly prove the $n$th order perturbation inequality when the first $(n-1)$ order perturbation inequalities are saturated. Using these results, we discuss the first $100$ order approximation of the gedanken experiments and find that the WCCC in Kerr black hole is valid under the higher-order approximation of perturbation. Our investigation implies that the WCCC might be strictly satisfied in Kerr black holes under the perturbation level.
}


\begin{document}

\maketitle

\section{Introduction}

General relativity, which describes gravitational interactions in terms of space-time curvature, has made great achievements in explaining many kinds of gravitational phenomena at various scales. Despite its success, however, general relativity still faces some theoretical challenges. One problem with general relativity is the existence of spacetime singularities in the solution of Einstein's equations. The Singularity theorem\cite{Penrose:1969pc} guarantees that the formation of singularities is a general feature of gravitational collapse. At the singularity, the physical quantities diverge. To ensure that there are no naked singularities in spacetime, Penrose proposed the weak cosmic censorship conjecture\cite{Penrose:1969pc,Wald:1997wa}.

Since the WCCC was proposed, it becomes one of the important conjectures in classical general relativity and still lacks general proof till now. In 1974, Wald first proposed a gedanken experiment\cite{Wald:1974wl} to test the validity of WCCC. They assumed a test particle absorbed by an extremal Kerr-Newman black hole and showed that the black hole cannot be over-spun or over-charged through this process under first-order approximation. After that, through dropping the test particle, Hubeny\cite{Hubeny:1998ga} considered the second-order approximation and found some possibilities of destruction. Then, violations of this type are found in many followup works\cite{deFelice:2001wj,Jacobson:2010iu,Chirco:2010rq,Saa:2011wq,Gao:2012ca}. However, Hubeny's method neglected some second-order effects such as the self-force and self-energy effects. In 2017, Sorce and Wald\cite{Sorce:2017dst} proposed a new version of the gedanken experiments. In this version, they considered a fully dynamic process of some matter field falling into a nearly extremal Kerr black hole. By assuming the matter fields satisfy the null energy condition, they used the Noether charge method\cite{Iyer:1994ys} to derive the first- and second-order perturbation inequalities
\ba\begin{aligned}
&\d M-\W_H\d J-\F_H\d Q\geq 0\,,\\
&\d^2 M-\W_H\d^2 J-\F_H\d^2 Q\geq -\frac{\k}{8\p} \d^2 A^\text{KN}_B
\end{aligned}\ea
to constrain the mass $M(\l)$, angular momentum $J(\l)$ and electric charge $Q(\l)$ of the geometry after perturbation. Using these results, they showed that the existence condition of the event horizon $h(\l)=M^2(\l)-Q^2(\l)-J^2(\l)/M^2(\l)$ can be reduced to
\ba\begin{aligned}\label{rr}
h(\l)\geq \left(\frac{(J^2-M^4)Q\d Q-2J M^2\d J}{M(M^4+J^2)}\l+M \e\right)^2\geq 0
\end{aligned}\ea
under the second-order approximation, in which $\e=r_h/M-1$ denotes the deviation to the extremal black hole. This result indicates that no violations can occur under the second-order approximation. Since then, this method has been extended to many other stationary black holes to test the validity of WCCC.

However, the discussion from Sorce and Wald only includes the first- and second-order approximation of the perturbation. According to their result \eq{rr}, we can see that there exists a second-order optimal condition in which the first two order perturbation inequalities are saturated and
\ba\begin{aligned}
\d Q=\frac{M^2[(M^4+J^2)\e-2J \d J\l]}{(M^4-J^2)Q\l}\,,
\end{aligned}\ea
such that $h(\l)=0$ under the second-order approximation and we cannot judge the validity of WCCC in this approximation. For strictness, it is necessary to consider high order approximation of the gedanken experiments. In 2020, one of our authors studied the high-order gedanken experiments for a nearly extremal Reissner-Nordstr\"{o}m black holes under the spherical perturbation\cite{Wang:2020vpn}, and found that the black hole cannot be overcharged in this process. However, one of the drawbacks of their method is that it is only suitable for the spherical symmetric perturbation process as well as the spherical spacetime. Astronomical observations show that most stars in the universe carry large angular momentum but small electric charge. Moreover, the real physical process in our universe is not spherical. Therefore, in this paper, we would like to extend the Sorce-Wald investigation into the high-order approximation in a nearly extremal Kerr black hole with any perturbation matter fields which satisfy the null energy condition and examine the WCCC in this process.

The remainder of this paper is organized as follows. In the next section, we review the Noether charge and variational identity in General Relativity. In section \ref{sec3}, we introduce the geometry of the Kerr black hole under the perturbation matter fields. In section \ref{sec4}, we perform the gedanken experiments proposed by Sorce and Wald\cite{Sorce:2017dst} to examine the WCCC under the high-order approximation of perturbation when the matter fields satisfy the null energy condition. Finally, we give a brief conclusion in section \ref{sec5}.

\section{Variational identity in Einstein gravity}
In this section, we review the Noether charge and variational identity in Einstein gravity with the Lagrangian four-form
\ba\begin{aligned}
\bm{L}_\text{grav}=\frac{\bm{\epsilon}}{16\p}R\,.
\end{aligned}\ea
Considering an one-parameter family with the parameter $\l$, the $k$th-order variation of the field $\f$ is defined by
\ba
\d^k \f =\left.\frac{d^k\f(\l)}{d\l^k}\right|_{\l=0}\,,
\ea
i.e., its $k$th derivative evaluated at $\l=0$. Taking the derivative of the Lagrangian, we have
\ba\label{fdeltalag}
	\frac{d \boldsymbol{L}_\text{grav}}{d\l} = \boldsymbol{E}_g^{ab} \frac{ d g_{ab}}{d\l} + d \boldsymbol{\Theta}\left(g, \frac{d g}{d\l}\right)\,,
\ea
where
\ba\begin{aligned}\label{theta}
\bm E_g^{ab}&=-\frac{\bm \epsilon}{16\p}G^{ab}\,,\\
  \boldsymbol{\Theta}_{abc} \left(g, \frac{d g}{d\l}\right)&= \frac{1}{16 \pi} \bm{\epsilon}_{dabc} g^{de} g^{fg} \left(\nabla_g \frac{d g_{ef}}{d\l} - \nabla_e \frac{d g_{fg}}{d\l} \right)
\end{aligned}\ea
are the on-shell equation of motion and the symplectic potential three-form separately. For any two-parameter family with parameters $\l_1$ and $\l_2$, we can define a symplectic current three form
\ba
	\bm\omega_{a b c}\left(g; \frac{\pd g}{\pd \l_1},\frac{\pd g}{\pd \l_2}\right) = \frac{1}{16 \pi} \bm{\epsilon}_{dabc} w^d\,,
\ea
in which
\ba\label{omega}
	w^a = P^{abcdef} \left(\frac{\pd g_{bc}}{\pd \l_2} \nabla_d \frac{\pd  g_{ef}}{\pd \l_1} - \frac{\pd g_{bc}}{\pd \l_1} \nabla_d \frac{\pd  g_{ef}}{\pd \l_2} \right)
\ea
with
\ba\begin{aligned}	
P^{abcdef} &= g^{ae} g^{fb} g^{cd} - \frac{1}{2} g^{ad} g^{be} g^{fc} - \frac{1}{2} g^{ab} g^{cd} g^{ef}- \frac{1}{2} g^{bc} g^{ae} g^{fd} + \frac{1}{2} g^{bc} g^{ad} g^{ef}.
\end{aligned}\ea
The Noether current three-form $\boldsymbol{J}_\z$ associated with the vector field $\z^a$ is defined by
\ba\begin{aligned}\label{Noethercurrent}
	\boldsymbol{J}_\z = \boldsymbol{\Theta} (g, \math{L}_\z g) - \z \cdot \boldsymbol{L}_\text{grav}\,.
\end{aligned}\ea
For another, the Noether current can also be expressed as\cite{Iyer:1994ys}
\ba\begin{aligned}\label{sjexp}
	\boldsymbol{J}_\z = \boldsymbol{C}_\z + d \boldsymbol{Q}_\z
\end{aligned}\ea
with the Noether charge two-form
\ba\begin{aligned}\label{chargegr}
	\left(\bm{Q}_\z \right)_{ab} = - \frac{1}{16 \pi} \bm{\epsilon}_{abcd} \nabla^c \z^d
\end{aligned}\ea
and the constraint
\ba\begin{aligned}\label{constraint}
	(\bm C_{\z})_{abc}=\frac{1}{8\p}\bm\epsilon_{eabc}\z^d G_d{}^e\,.
\end{aligned}\ea
The derivative of the identity
\ba\begin{aligned}
  \boldsymbol{\Theta} (g, \math{L}_\z g) - \z \cdot \boldsymbol{L}_\text{grav}=\boldsymbol{C}_\z + d \boldsymbol{Q}_\z
\end{aligned}\ea
gives the variational identity
\ba\begin{aligned}\label{variation1app}
  \quad d&\left[\frac{d \boldsymbol{Q}_\z}{d\l} - \z \cdot \boldsymbol{\Theta} \left(g, \frac{d g}{d\l} \right)  \right]=\bm\omega\left(g;\frac{d g}{d\l},\mathcal{L}_\z g\right)-\z \cdot \boldsymbol{E}_g^{ab} \frac{d g_{ab}}{d\l} - \frac{d\boldsymbol{C}_\z}{d\l}\,.
\end{aligned}\ea
It is necessary to note that the above identity holds for any spacetime configuration even it does not satisfy the on-shell equation of motion $G_{ab}=0$.

\section{Perturbed geometry of Kerr black hole}\label{sec3}
In this section, we would like to introduce the perturbation geometry of the Kerr black hole and introduce the process that some perturbation matters fall into a nearly extremal Kerr black hole. In this case, the full Lagrangian is given by
\ba\begin{aligned}
\bm L=\frac{\bm \epsilon}{16\pi}R+\bm L_\text{mt}\,,
\end{aligned}\ea
where we use $\bm L_\text{mt}$ to denote the Lagrangian of the perturbation matter fields. From the variation of the Lagrangian, we can find the equation of motion is
\ba
G_{ab}=T_{ab}\,,
\ea
where $T_{ab}$ is the stress-energy tensor of the perturbation matter field. Now, we consider a perturbation to the Kerr black hole, i.e., we have $T_{ab}=0$ in the background geometry. The metric of the Kerr solution in the Boyer-Lindquist coordinate system can be written as
\ba\begin{aligned}\label{backgroundmetric}
  ds^2=&-\left[1-\frac{2Mr}{\rho^2}\right]dt^2+\frac{\rho^2}{\Delta}dr^2+\rho^2 d\theta^2\\
  &+\left[(r^2+a^2)^2-\Delta a^2 \sin^2 \theta\right]\frac{\sin^2\theta}{\rho^2}d\j^2-\frac{4aMr\sin^2\theta}{\rho^2}dt d\j
\end{aligned}\ea
with
\ba\begin{aligned}
&a=J/M\,,\quad\quad \rho^2=r^2+a^2\cos^2\theta\,,\quad\quad\Delta(r)=r^2-2Mr+a^2\,,
\end{aligned}\ea
in which $M$ and $J$ are the mass and angular momentum of the spacetime and they are defined by
\ba\begin{aligned}
\delta M=\int_{\infty}\delta\bm Q_{t}-t\cdot\bm\Theta(g,\delta g)\,,\quad\quad
\delta J=\int_{\infty}\delta\bm Q_{\j}\,.
\end{aligned}\ea
Here
\ba t^a=\left(\frac{\partial}{\partial t}\right)^a\,,\quad \j^a=\left(\frac{\partial}{\partial \j}\right)^a
\ea
are the killing vectors related to the time evolution and axial symmetry at asymptotically infinity, individually.

The horizon of the black hole is given by the positive root of $\Delta(r_h)=0$, i.e.,
\ba\begin{aligned}
r_h=M+\sqrt{M^2-J^2/M^2}\,.
\end{aligned}\ea
That is to say, when $M^2\geq J^2/M^2$, the solution describes a black hole geometry; when $M^2< J^2/M^2$, the solution describes a naked singularity. Especially, when $M^4=J^2$, it describes an extremal black hole.

For the black hole case, the area, angular velocity, and surface gravity of the event horizon is given by
\ba\begin{aligned}\label{AWk}
A=4\p \left(r_h^2+J^2/M^2\right)\,,\quad\quad\W_H=\frac{J/M}{r_h^2+J^2/M^2}\,,\quad\quad \k=\frac{M^2-J^2/M^2}{2M r_h}\,.
\end{aligned}\ea

\begin{figure}
\centering
\includegraphics[width=0.8\textwidth]{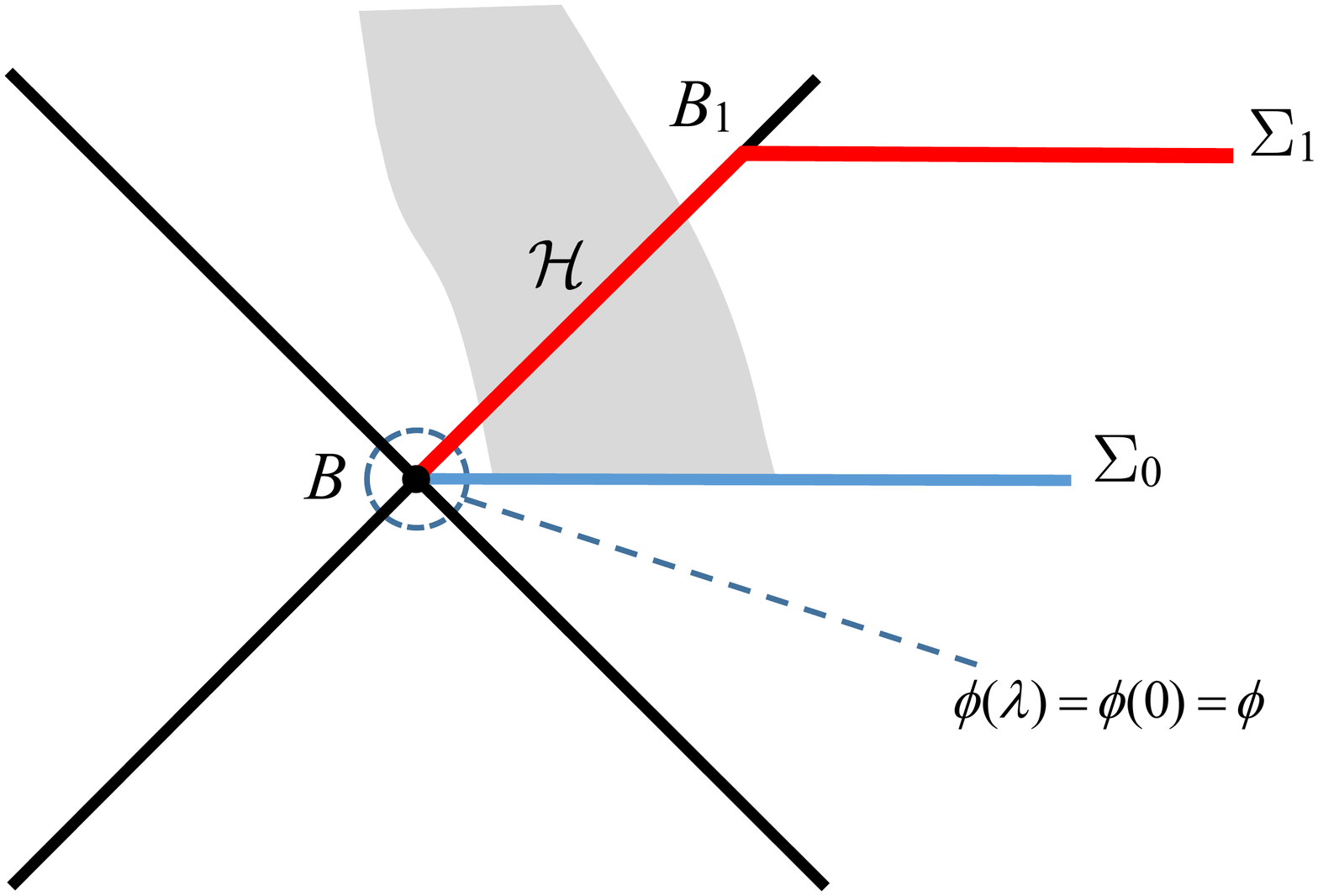}
\caption{Plot showing the perturbed geometry of the Kerr black hole and the choice of the hypersurface $\S=\math{H}\cup \S_1$. }\label{fig1}
\end{figure}

To test the WCCC, we consider a perturbation process that some collision matters fall into the future horizon during a finite time\cite{Sorce:2017dst}, i.e., the perturbation matter fields vanish at sufficiently late times as well as some early time before perturbation (See \fig{fig1}). Let $\f(\l)$ to be a one-parameter family in the configuration space, and each element $\phi(\lambda)$ in this family represents a perturbation process and $\f=\phi(0)$ describes the background geometry, which can be described by the Kerr metric with the mass $M$ and angular momentum $J$. Here we introduce $\f$ to denote the collection of the metric $g_{ab}$ and the perturbation matter fields. We assume that the perturbation vanishes at early times, i.e., the spacetime before the perturbation is the same as the background, and we also assume that the spacetime satisfies the stability condition\cite{Sorce:2017dst, Hollands:2012sf}, which means that the late-time geometry can also be described by the Kerr metric with different mass $M(\l)$ and angular momentum $J(\l)$ and therefore we have $\bm E_g^{ab}(\l)=\bm{C}_\z(\l)=0$ at late times.

Under the stability condition, examining the WCCC is equivalent to checking whether the late-time geometry also describes a black hole. Therefore, we can define a function
\ba\begin{aligned}\label{blackening}
h(\lambda)=M(\lambda)^4-J(\lambda)^2
\end{aligned}\ea
If $h(\lambda)\geq 0$, there exists an event horizon after perturbation and the WCCC is satisfied in this process. If $h(\lambda)<0$, the black hole is destroyed and the WCCC is violated by the perturbation process.

In this paper, we would like to consider some perturbation matter falling into a nearly extremal Kerr black hole. Without loss of generality, we set $M=1$ for the mass of the background black hole. For the nearly extremal black holes, we can define a small quantity $\epsilon$ given by
\ba\begin{aligned}\label{extremal}
J=\sqrt{1-\epsilon^2}
\end{aligned}\ea
to describe the deviation from the extremal black hole. To make the black hole easier to be destroyed, this parameter is chosen to be in the same order with $\lambda$ \cite{Sorce:2017dst}. Expanding $h(\l)$ at $\l=0$, we have
\ba\begin{aligned}\label{hlambda}
h(\lambda)=&1-J^2+\lambda\left(4\d M-2J\d J\right)+\lambda^2\left(6\d M^2+2\d^2 M-\d J^2-J\d^2 J\right)\\
&+\frac{\lambda^3}{3}(12\d M^3+18\d M\d^2M+2\d^3M-3\d J\d^2 J-J\d^3J)\\
&+\frac{\lambda^4}{12}\left(12\d M^4+72\d M\d^2M+18\d^2M^2+24\d M\d^3M\right.\\
&\left.+2\d^4M-3\d^2J^2-4\d J\d^3J-J\d^4J\right)+\math{O}(\l^4)\,.
\end{aligned}\ea
Therefore, the key point to check the signature of $h(\l)$ under the perturbation process is to get the constraint of the variational quantities $\d^k M$ and $\d^k J$ from the null energy condition of the perturbation matter fields.

\section{Gedanken experiments to overspin the black holes}\label{sec4}

Next, we would like to derive the perturbation inequality by assuming the perturbation matter field satisfies the null energy condition and examining whether the horizon exists after the perturbation. The null energy condition of the perturbation matter fields demand
\ba\begin{aligned}
T_{ab}(\l)l^a l^b \geq 0\,,
\end{aligned}\ea
in which $l^a$ is any future-directed null vector in the configuration $\f(\l)$.

With a same setup as Ref. \cite{Sorce:2017dst}, we choose a hypersurface $\Sigma=\math{H}\cup \Sigma_1$ as shown in \fig{fig1} satisfying the following conditions: (1) $\math{H}$ is a null hypersurface which starts from the bifurcation surface $B$ before the perturbation and finally ends at a very late-time cross-section $B_1$ where all the perturbation matter fields vanish. (2) $\Sigma_1$ is a spacelike hypersurface that starts from $B_1$ and extends to infinity. (4) $\math{H}$ is the Killing horizon after perturbation and therefore it is the event horizon in the background geometry. Let $\x^a$ be the null generator of the hypersurface $\math{H}$. In the background geometry, it becomes the Killing vector of the event horizon, i.e.,
\ba\begin{aligned}
\xi^a=t^a+\Omega_H\j^a
\end{aligned}\ea
On the background, $\xi^a$ is the Killing vector field of the background horizon and the tangent vector of $\math{H}$. Choose a gauge to make the hypersurface $\S$ and the vector field $\x^a$ fixed under variation. Replacing $\z^a$ by $\x^a$, integration of the variational identity \eq{variation1app} on $\S$ gives
\ba\begin{aligned}\label{varid}
  \int_\inf\left[\frac{d \boldsymbol{Q}_\x}{d\l} - \x \cdot \boldsymbol{\Theta} \left(g, \frac{d g}{d\l} \right)  \right]&=\int_\S\bm\omega\left(g;\frac{d g}{d\l},\mathcal{L}_\x g\right) -\int_{\math{H}}\frac{d\boldsymbol{C}_\x}{d\l}\,,
\end{aligned}\ea
where we have used that the assumptions that the perturbation vanishes on $B$, i.e., $d g(\l)/d\l=0$ on $B$, and $\x^a$ is a tangent vector on $\math{H}$ and the perturbation matter fields vanish at late times, i.e., we have $\bm E_g^{ab}(\l)=\bm C_\x(\l)=0$ on $\S_1$. In the following, we would like to perform this identity to get the constraint from the null energy condition of the perturbation matter fields.

\subsection{First-order approximation of perturbation}

With a calculation same with Sorce and Wald, after evaluating the identity \eq{varid} at $\l=0$ and using the stationary condition of the background spacetime $\math{L}_\x g_{ab}=0$, we further obtain
\ba\begin{aligned}\label{varid1}
\delta M-\Omega_H\delta J=-\int_{\math{H}}\delta\bm C_{\xi}=\int_{\math{H}}\tilde{\bm{\epsilon}}\d T_{ab}k^a \x^b \,,
\end{aligned}\ea
where $\tilde{\bm{\epsilon}}$ is the induced volume element of $\math{H}$ which is defined by $\bm{\epsilon}_{ebcd}=-4 k_{[e}\tilde{\bm{\epsilon}}_{bcd]}$ and $k^a\propto \x^a$ is the future-directed normal vector of $\math{H}$. Considering that $T_{ab}=0$ in the background geometry, the null energy condition under the first-order approximation implies that $\d T_{ab} \x^a \x^b\geq 0$. Then, we can get the first-order perturbation inequality
\ba\begin{aligned}\label{firstineq}
\delta M-\Omega_H \delta J\geq 0
\end{aligned}\ea
with the angular momentum
\ba\begin{aligned}
\Omega_H=\frac{1}{2}\sqrt{\frac{1-\epsilon}{1+\epsilon}}\,.
\end{aligned}\ea
Together with Eq. \eq{extremal} and Eq. \eq{hlambda}, we have
\ba\begin{aligned}
h(\lambda)=\lambda(4\d M-2\d J)\geq 0
\end{aligned}\ea
under the first-order approximation and therefore the black hole cannot be overspun under the first-order approximation of perturbation. However, this does not mean that the WCCC is valid in the perturbation process since there exists a first-order optimal condition $\delta M=\Omega_H \delta J$ such that $h(\l)=0$ under the first-order approximation and thus the signature of $h(\l)$ is determined by the higher-order approximation. That is to say, to strictly test the WCCC in the perturbation process, we need to consider the second- or even higher-order approximations.

Finally, we would like to discuss the first-order optimal condition. From Eq. \eq{varid1}, we can see that this condition implies
\ba\begin{aligned}\label{firstop1}
\d T_{ab}\x^a \x^b=0\,.
\end{aligned}\ea
Let $u$ be the parameter of $\x^a$, the expansion and shear of the null hypersurface $\math{H}$ are defined by
\ba\begin{aligned}
\vartheta =\frac{1}{2}\g^{ab}\math{L}_\x\g_{ab}\,,\quad\quad \s_{ab}=\frac{1}{2} \math{L}_\x \g_{ab}-\frac{\vartheta}{2}\g_{ab}\,,
\end{aligned}\ea
where $\g_{ab}$ is the induced metric of the cross-section on $\math{H}$.  Considering the Raychaudhuri equation is given by
\ba\begin{aligned}\label{Ray}
  \frac{d\vartheta}{d u}=-\frac{1}{2}\vartheta^2-\sigma_{ab}\sigma^{ab}-\xi^a\xi^b T_{ab}+\k \vartheta\,,
\end{aligned}\ea
the optimal condition gives
\ba\begin{aligned}\label{firstop2}
\frac{d\d\vartheta}{d u}=\k \d \vartheta\,,
\end{aligned}\ea
where we used the fact that $\vartheta =\s_{ab}=0$ in the background geometry. Combing with the assumption that the perturbation vanishes (i.e., $\d\vartheta =0$) in the region near the bifurcation surface, the above result also implies that
\ba\label{firstop3}
\d \vartheta=0
\ea
on $\math{H}$.

\subsection{Second-order approximation of peturbation}

Next, we will derive the second-order perturbation inequality under the first-order optimal condition. Taking a derivative of Eq. \eq{varid}, a same calculation as Ref. \cite{Sorce:2017dst} gives
\ba\begin{aligned}\label{variation2}
  \delta^2 M-\Omega_H\delta^2 J=\mathcal{E}_{\Sigma}(g,\delta g)-\int_{\math{H}}\delta^2\bm C_{\xi}\,,
\end{aligned}\ea
where we define the canonical energy as
\ba\begin{aligned}
 \mathcal{E}_{\Sigma}(g,\delta g)=\int_{\Sigma} \bm\omega(g;\delta g,\mathcal{L}_{\xi}\delta g)\,.
\end{aligned}\ea
Using the first-order optimal condition $\d T_{ab}\x^a \x^b=0$, we have
\ba\begin{aligned}
\int_\math{H}\delta^2\bm C_{\xi}= -\int_\math{H}\tilde{\bm \epsilon}k^a\x^b\delta^2T_{ab}\leq 0\,.
\end{aligned}\ea
Thus, we can obtain
\ba\begin{aligned}\label{secondorderE}
\delta^2 M-\Omega_H\delta^2 J\geq \mathcal{E}_{\Sigma}(g,\delta g)
\end{aligned}\ea
The canonical energy can be divided into two parts:
\ba\begin{aligned}
\mathcal{E}_{\Sigma}=\mathcal{E}_\math{H}+\mathcal{E}_{\Sigma_1}=\int_\math{H}\bm\omega(g;\delta g,\mathcal{L}_{\xi}\delta g)+\int_{\Sigma_1}\bm\omega(g;\delta g,\mathcal{L}_{\xi}\delta g)
\end{aligned}\ea
Next, we refer to the method by Sorce and Wald to calculate $\mathcal{E}_{\Sigma_1}(\phi,\delta\phi)$. We consider another one-parameter family $\phi_{(2)}(\a)$ in which every element is given by the Kerr spacetime with the mass and angular momentum
\ba\begin{aligned}\label{MK1}
M_{(2)}(\a)=M+\a \delta M\,,\quad\quad J_{(2)}(\a)=J+\a \delta J\,.
\end{aligned}\ea
Here $\d M$ and $\d J$ are chosen to be agree with the quantities evaluated by the one-parameter family $\f(\l)$. Since there is only the first-order variation of $\f(\l)$ in $\mathcal{E}_{\Sigma_1}(\phi,\delta\phi)$, we have
\ba\begin{aligned}
\mathcal{E}_{\Sigma_1}(\phi,\delta\phi)=\mathcal{E}_{\Sigma_1}(\phi_{(2)},\delta\phi_{(2)})\,.
\end{aligned}\ea
In this family, considering that $\delta^2 M_{(2)}=\delta^2 J_{(2)}=0$, we can obtain \cite{Sorce:2017dst,Hollands:2012sf}
\ba\begin{aligned}
  \mathcal{E}_{\Sigma_1}(g,\delta g)=-\frac{\kappa}{8\pi}\delta^2 A_{(2)}\,,
\end{aligned}\ea
in which $A_{(2)}(\a)$ is the area of the bifurcation surface $B$ in the configuration $\f_{(2)}(\a)$. Through a straightforward calculation, we can obtain
\ba\begin{aligned}
\delta^2A_{(2)}=-4\pi\left(1+\frac{1}{\epsilon}\right)\delta J^2\,,
\end{aligned}\ea
where we used the first-order optimal condition $\delta M-\Omega\delta J=0$ to simplify.

Then, we would like to evaluate the horizon part of the canonical energy. To do this, we introduce the Gaussian null coordinate system of the hypersurface $\math{H}$. Let the null tangent vector $\xi^a=(\partial/\partial u)^a$ be the first coordinate basis. Another coordinate basis $(\partial/\partial z)^a$ is given by another null vector and satisfies
\ba\begin{aligned}
\left.\left(\frac{\partial}{\partial u}\right)^a\left(\frac{\partial}{\partial z}\right)_a\right|_\math{H}=1\,.
\end{aligned}\ea
Then, the metric near the null hypersurface $\math{H}$ can be expressed as
\ba\begin{aligned}\label{guassiannull}
ds^2=2(dz-z\alpha du-z\b_i d\q^i)du+\g_{ij} d\q^id\q^j\,.
\end{aligned}\ea
Here $z=0$ gives the location of $\math{H}$. It is necessary to mention that this coordinate system is compatible with the gauge choice in previous calculations, i.e., fixing the coordinate in the variation also makes $\x^a$ and $\math{H}$ fix.

Using the metric, we can find $\d g_{ab}=\d \g_{ab}$ on $\math{H}$. Using the definition of the expansion, the first-order optimal condition $\d\vartheta =0$ implies
\ba\label{optimal1}
g^{ab}\math{L}_\x \d g_{ab}=\g^{ab}\math{L}_\x \d \g_{ab}=0\,,\quad\text{and} \quad \delta\sigma_{ab}=\frac{1}{2}\mathcal{L}_{\xi}\delta\gamma_{ab}
\ea
on $\math{H}$. Considering the assumption that the perturbation vanishes at early times, the first condition of the above expressions also implies
\ba\begin{aligned}
g^{ab}\d g_{ab}=\g^{ab}\d \g_{ab}=0
\end{aligned}\ea
on $\math{H}$.

According to Eq. \eqref{omega}, $\mathcal{E}_\math{H}(g,\delta g)$ can be written as
\ba\begin{aligned}\label{EH}
  \mathcal{E}_\math{H}(g,\delta g)&=-\frac{1}{16 \pi}\int_\math{H}\bar{\bm\epsilon}\xi_a P^{abcdef}\left(\mathcal{L}_{\xi}\delta g_{bc}\nabla_d\delta g_{ef}-\delta g_{bc}\nabla_d \mathcal{L}_{\xi}\delta g_{ef}\right)\\
  &=-\frac{1}{8\pi}\int_\math{H}\bar{\bm\epsilon}\bar{P}^{zbcdef}\mathcal{L}_{\xi}\delta g_{bc}\nabla_d\delta g_{ef}+\frac{1}{16\pi}\int_\math{H}\mathcal{L}_{\xi}(\bar{\bm\epsilon}\bar{P}^{zbcdef}\delta g_{bc}\nabla_d \delta g_{ef})\,,
\end{aligned}\ea
where we have defined
\ba\begin{aligned}
\bar{P}^{abcdef}=g^{ae} g^{fb} g^{cd} - \frac{1}{2} g^{ad} g^{be} g^{fc} - \frac{1}{2} g^{ab} g^{cd} g^{ef}\,,
\end{aligned}\ea
and used the first-order optimal condition $g^{ab}\math{L}_\x \d g_{ab}=0$ on $\math{H}$. Here we introduce the notations
\ba\begin{aligned}
\bar{\bm {\epsilon}}=\sqrt{\g} du\wedge d\q^1\wedge d\q^2\,,\quad \quad\hat{\bm {\epsilon}}=\sqrt{\g}d\q^1\wedge d\q^2\,.
\end{aligned}\ea
With a straightforward calculation using the Gaussian null coordinate system, it is easy to find
\ba\begin{aligned}\label{Pabcdef}
\bar{P}^z{}_{ij}{}^{def}\nabla_d\delta g_{ef}=-\frac{1}{2}\delta\gamma_{ij;u}=-\delta\sigma_{ij}+\delta\gamma_{k(i}\Gamma^k{}_{j)u}\,.
\end{aligned}\ea
On background, we have $\Gamma^i{}_{ju}=0$. Besides, according to the assumption by Sorce and Wald\cite{Sorce:2017dst}, the perturbation is physically stationary on $B_1$, which means $\delta\sigma_{ab}=0$ on $B_1$. Together with the first-order optimal condition \eqref{optimal1}, we have
\ba\begin{aligned}
\mathcal{E}_H(\phi,\delta\phi)&=\frac{1}{4\pi}\int_H\bar{\bm\epsilon}\delta\sigma_{ab}\delta\sigma^{ab}+\frac{1}{16\pi}\int_{B_1}\hat{\bm\epsilon}\delta g^{ab}\delta\sigma_{ab}\\
&=\frac{1}{4\pi}\int_H\bar{\bm\epsilon}\delta\sigma_{ab}\delta\sigma^{ab}\geq 0\,.
\end{aligned}\ea
Summarizing the above results, the second-order perturbation inequality gives
\ba\begin{aligned}\label{secondorderE}
\delta^2 M-\Omega_H\delta^2 J\geq -\frac{\k}{8\p}\d^2 A_{(2)}=\frac{\d J^2}{4}\,.
\end{aligned}\ea
where we used
\ba
\k=\frac{\epsilon}{2+2\epsilon}
\ea
and the first-order optimal condition.

After considering the first-order optimal condition and the second-order perturbation inequality, we can get
\ba\begin{aligned}
h(\lambda)\geq(\epsilon-\lambda\delta J)^2\geq 0
\end{aligned}\ea
under the second-order approximation of perturbation, i.e., we neglected the higher-order terms $\mathcal{O}(\epsilon^3,\lambda^3,\epsilon\lambda^2,\dots)$. This result implies that the nearly extremal Kerr black hole cannot be overspun in the second-order approximation of perturbation. Same to the first-order case, there also exists a second-order optimal condition in which the first two perturbation inequalities are saturated and
\ba
\d J=\epsilon/\l\,,
\ea
in which we cannot determine the validity of the WCCC in the second-order approximation. Therefore, we need to extend the discussion into the third- or even higher-order approximation.

With the same discussion as the first-order case, it is not hard to verify that the second-order optimal condition will give the additional conditions
\ba\begin{aligned}\label{optimalT2}
&\delta^2 T_{ab}\xi^a\xi^b=0\,,\quad\quad \delta\sigma_{ab}=0\,.
\end{aligned}\ea
Together with $\d \vartheta=0$, these also gives
\ba\begin{aligned}
\math{L}_{\x}\d \g_{ab}=\d \g_{ab}=0
\end{aligned}\ea
on $\math{H}$. Using the above conditions and combing with the first-order optimal condition, the second variation of the Raychaudhuri equation gives
\ba
\frac{d\d^2\vartheta}{d u}=\k \d^2\vartheta\,.
\ea
Considering that $\d^2\vartheta =0$ on $\math{H}$ at early times, this implies that $\d^2\vartheta=0$ on the whole $\math{H}$.

\subsection{Third-order approximation of perturbation}

In this subsection, we extend the discussion into the third-order approximation. Taking two variation to Eq. \eqref{varid}, similar calculation can give
\ba\begin{aligned}
  \delta^3 M-\Omega_H\delta^3 J&=\delta\mathcal{E}_\math{H}(g,\delta g)+\delta\mathcal{E}_{\Sigma_1}(g,\delta g)+\int_H\tilde{\bm\epsilon}k^a\xi^b\delta^3T_{ab}\\
&\geq\delta\mathcal{E}_\math{H}(g,\delta g)+\delta\mathcal{E}_{\Sigma_1}(g,\delta g)\,,
\end{aligned}\ea
where we used the null energy condition $\d^3 T_{ab}\x^a\x^b\geq 0$ of the perturbation matter field under the third-order approximation in the second-order optimal condition.

First, we evaluate the canonical energy contributed by $\S_1$. Considering that $\d \math{E}_{\S_1}(g, \d g)$ only depends on the first two order variation of $g_{ab}$. We can also calculate it by introducing a one-parameter family $\f_{(3)}(\a)$ in which any element describes a Kerr solution with the mass and angular momentum
\ba\begin{aligned}
  &M_{(3)}(\a)=M+\a \delta M+\frac{1}{2}\a^2\delta^2 M\,,\\
  &J_{(3)}(\a)=J+\a \delta J+\frac{1}{2}\a^2\delta^2 J\,,
\end{aligned}\ea
where $\d M, \d^2 M, \d J$ and $\d^2 J$ are chosen to agree with the variations from the configuration $\f(\l)$ under the first-two order approximation. With a similar calculation, we can also get
\ba\begin{aligned}
  \delta\mathcal{E}_{\Sigma_1}(g,\delta g)&=-\frac{\kappa}{8\pi}\delta^3 A_{(3)}=\frac{3\delta J}{8}\left(2\delta^2J-\sqrt{\frac{1-\epsilon}{1+\epsilon}}\delta J^2\right)\,.
\end{aligned}\ea

Finally, we turn to evaluate the canonical energy from the hypersurface $\math{H}$.  From the second-order optimal condition, we also have
\ba\begin{aligned}\label{optimalsecondorder}
&g^{ab}\mathcal{L}_{\xi}\delta^2g_{ab}=g^{ab}\delta^2g_{ab}=0\,,\quad\quad \delta^2\sigma_{ab}=\frac{1}{2}\mathcal{L}_{\xi}\delta^2\gamma_{ab}\,,\quad\quad \d g_{ab}=\d \g_{ab}=\delta\sigma_{ab}=0
\end{aligned}\ea
on $\math{H}$. In addition, since $\xi^a$ is killing vector on the background, it is easy to check
\ba\begin{aligned}
\delta(\nabla_d\mathcal{L}_{\xi}\delta g_{ef})=\nabla_d(\mathcal{L}_{\xi}\delta^2 g_{ef})=\mathcal{L}_{\xi}(\nabla_d\delta^2 g_{ef})
\end{aligned}\ea
According to Eq. \eqref{omega}, $\d \mathcal{E}_\math{H}(g,\delta g)$ can be written as
\ba\begin{aligned}\label{EH2}
  \delta\mathcal{E}_\math{H}(g,\delta g)
  &=-\frac{1}{8\pi}\int_\math{H}\bar{\bm\epsilon}\bar{P}^{zbcdef}\mathcal{L}_{\xi}\delta^2 g_{bc}\nabla_d\delta g_{ef}+\frac{1}{16\pi}\int_\math{H}\mathcal{L}_{\xi}(\bar{\bm\epsilon}\bar{P}^{zbcdef}\delta g_{bc}\nabla_d \delta^2 g_{ef})\\
  &=\frac{1}{4\pi}\int_\math{H}\bar{\bm\epsilon}\delta^2\sigma_{ab}\delta\sigma^{ab}+\frac{1}{16\pi}\int_{B_1}\hat{\bm\epsilon}\delta g^{ab}\delta^2\sigma_{ab}=0
\end{aligned}\ea
where we used the second-order optimal condition Eq. \eq{optimalsecondorder}.

Summarizing the above results, the third-order perturbation inequality can be shown as
\ba\begin{aligned}
\delta^3 M-\Omega_H\delta^3 J\geq -\frac{\kappa}{8\pi}\delta^3 A_{(3)}\,.
\end{aligned}\ea

After taking these conditions into account and together with the second-order optimal condition, it is not hard to show
\ba\begin{aligned}
h(\l)\geq 0
\end{aligned}\ea
under the third-order approximation of perturbation, i.e., we neglect the higher-order terms $\mathcal{O}(\epsilon^4,\lambda^4, \l \epsilon^3\dots)$. This result shows that the WCCC cannot be violated under the fourth-order approximation. Similarly, there also exists a third-order optimal condition leading us to consider the higher-order approximations, i.e., the second-order optimal condition is satisfied and the third-order perturbation inequality is saturated.

Under the third-order optimal condition, a similar calculation gives the following additional conditions:
\ba\begin{aligned}
\d^3 T_{ab}\x^a\x^b=0\,,\quad\quad \d^3\vartheta =0
\end{aligned}\ea
on $\math{H}$

\subsection{Fourth-order approximation of perturbation}
 Taking two variations to Eq. \eqref{varid} and considering the null energy condition $\d^4 T_{ab}\x^a\x^b\geq 0$ of the perturbation matter fields under the third-order optimal condition, the similar calculation can give
\ba\begin{aligned}
  \delta^4 M-\Omega_H\delta^4 J&=\delta^2\mathcal{E}_\math{H}(g,\delta g)+\delta^2\mathcal{E}_{\Sigma_1}(g,\delta g)+\int_\math{H}\tilde{\bm\epsilon}k^a\xi^b\delta^4T_{ab}\\
&\geq\delta^2\mathcal{E}_\math{H}(g,\delta g)+\delta^2\mathcal{E}_{\Sigma_1}(g,\delta g)\,.
\end{aligned}\ea

For the canonical energy contributed by $\S_1$. Similarly, it can be calculated in the one-parameter family $\f_{(4)}(\a)$ constructed by the Kerr spacetime with
\ba\begin{aligned}
  &M_{(4)}(\a)=M+\a \delta M+\frac{1}{2}\a^2\delta^2 M+\frac{1}{3!}\a^3\delta^3 M\,,\\
  &J_{(4)}(\a)=J+\a \delta J+\frac{1}{2}\a^2\delta^2 J+\frac{1}{3!}\a^3\delta^3 J\,.
\end{aligned}\ea
Then, we can get
\ba\begin{aligned}
  \delta^2\mathcal{E}_{\Sigma_1}(g,\delta g)&=-\frac{\kappa}{8\pi}\delta^4 A_{(4)}\\
  &=\d J \d^3J+\frac{3}{4}\d^2 J^2-\frac{9\sqrt{1-\epsilon}}{4\sqrt{1+\epsilon}}\d J^2\d^2 J+\frac{3(3-5\epsilon)}{16(1+\epsilon)}\d J^4\,.
\end{aligned}\ea

The third-order optimal condition gives
\ba\begin{aligned}
&g^{ab}\mathcal{L}_{\xi}\delta^3g_{ab}=g^{ab}\delta^3g_{ab}=0
\end{aligned}\ea
on $\math{H}$, where we have used the second-order optimal condition $\d g_{ab}=0$ on $\math{H}$.
Then, $\d^2 \mathcal{E}_\math{H}(g,\delta g)$ can be shown as
\ba\begin{aligned}\label{31}
\delta^{2}\mathcal{E}_\math{H}( g,\delta g)=-\frac{1}{8 \pi}\int_\math{H}\bar{\bm\epsilon} \left[\mathcal{L}_{\xi}\delta^{2} \gamma_{bc} \delta(\bar{P}^{zbcdef}\nabla_d\delta g_{ef})-\delta^2\gamma_{bc} \delta(\bar{P}^{zbcdef}\nabla_d\mathcal{L}_{\xi}\delta g_{ef})\right]\,.\quad
\end{aligned}\ea
We calculate the variation term in the first term and find
\ba\begin{aligned}\label{1}
  \delta(\bar{P}^{z}{}_{ij}{}^{def}\nabla_d\delta g_{ef})&=\delta(-\delta\sigma_{ij}+\delta\gamma_{k(i}\Gamma^k{}_{j)u})\\
  &=-\delta^{2}\sigma_{ij}+\sum_{m=0}^{1}\delta^{m+1}\gamma_{k(i}\delta^{1-m}\Gamma^k{}_{j)u}\\
  &=-\delta^2\sigma_{ij}\,,
\end{aligned}\ea
as well as
\ba\begin{aligned}\label{2}
  \delta(\bar{P}^{z}{}_{ij}{}^{def} \nabla_d \mathcal{L}_{\xi}\delta g_{ef})&=\bar{P}^{z}{}_{ij}{}^{def}\mathcal{L}_{\xi}\nabla_d\delta^{2}g_{ef}=\mathcal{L}_{\xi}(\bar{P}^{z}{}_{ij}{}^{def}\nabla_d\delta^{2}g_{ef})\\
  &=\mathcal{L}_{\xi}(-\delta^{2}\sigma_{ij}+\delta^{2}\gamma_{k(i}\Gamma^k{}_{j)u})=-\mathcal{L}_{\xi}\delta^2\sigma_{ij}\,,
\end{aligned}\ea
where we have used the third-optimal condition, $\d g_{ab}=\d \g_{ab}=0$ and $\G^i{}_{iu}=0$ on $\math{H}$. Using the above results and considering the stability condition $\d^2\s_{ab}=0$ on $B_1$, we can further obtain
\ba\begin{aligned}
  \delta^{2}\mathcal{E}_\math{H}( g,\delta g)
  &=\frac{1}{4 \pi}\int_\math{H}\bar{\bm\epsilon} \mathcal{L}_{\xi}\delta^{2} \gamma_{bc} \delta^{2}\sigma^{bc}+\frac{1}{8\pi}\int_{B_1}\hat{\bm\epsilon}\delta^{2} \gamma_{bc}\delta^{2}\sigma^{bc}\\
  &=\frac{1}{2 \pi}\int_{H}\bar{\bm\epsilon} \delta^{2}\sigma_{bc} \delta^{2}\sigma^{bc}\geq 0\,,
\end{aligned}\ea
Combing these results, the fourth-order perturbation inequality can be given by
\ba\begin{aligned}
\delta^4 M-\Omega_H\delta^4 J\geq -\frac{\kappa}{8\pi}\delta^4 A_{(4)}\,.
\end{aligned}\ea

After a straightforward calculation, we find
\ba\begin{aligned}
h(\lambda)\geq\frac{1}{4}(\epsilon^2-\lambda^2\delta^2 J)^2\geq 0
\end{aligned}\ea
under the fourth-order approximation of $\lambda$ and $\epsilon$. Therefore the WCCC cannot be violated under the fourth-order approximation. Similarly, there exits a fourth-order optimal condition in which $\d^2 J=\epsilon^2/\l^2$, fourth-order perturbation inequality is saturated, and the third-order optimal condition is satisfied, such that the fourth-order approximation cannot judge the signature of $h(\l)$.

Similar discussions can show that the fourth-order optimal condition can also give
\ba\begin{aligned}\label{optimalT2}
&\delta^4 T_{ab}\xi^a\xi^b=0\,,\quad\quad \delta^2\sigma_{ab}=0\,.
\end{aligned}\ea
These imply
\ba\begin{aligned}
\d^4\vartheta=0\,,\quad\quad\d^2 \g_{ab}=\d^2 g_{ab}=0\,.
\end{aligned}\ea

\subsection{$k$th-order approximation of perturbation}

\subsubsection{$k$th-order perturbation inequality}
In this subsection, we would like to discuss the gedanken experiment under the higher-order perturbation inequality. By summarizing the previous results, we would like to derive the following $n$th-order perturbation inequality
\ba\begin{aligned}\label{ineqn}
\d^n M-\W_H\d^n J\geq -\frac{\k}{8\p} \d^n A_{(n)}
\end{aligned}\ea
when the first $(n-1)$th order perturbation inequalities are saturated, and the saturation of the first $n$th order perturbation inequalities give
\ba\begin{aligned}\label{optimaln}
&\delta^{i} T_{ab}\xi^a\xi^b=0\,,\quad\forall\, i\leq n\,,\\
&\delta^i\vartheta=\frac{1}{2}\gamma^{ab}\mathcal{L}_{\xi}\delta^i\gamma_{ab}=0\,,\quad\forall\, i\leq n\,,\\
&\delta^j g_{ab}=\delta^j\gamma_{ab}=0\,, \quad\forall\, 1\leq j\leq [n/2]\,,
\end{aligned}\ea
on the hypersurface $\math{H}$, where $[n/2]$ denote the integer part of $n/2$. Here $A_{(n)}(\a)$ is the area of bifurcation surface $B$ in the one-parameter family $\f_{(n)}(\a)$ constructed by the Kerr solution with the parameters
\ba\begin{aligned}
M_{(n)}(\a)=\sum _{i=0}^{n-1} \frac{\alpha^i \delta^i M}{i!}\,,\quad\quad J_{(n)}(\a)=\sum _{i=0}^{n-1} \frac{\alpha^i \delta^i J}{i!}\,.
\end{aligned}\ea
Next, we would like to perform the mathematical induction to prove the above result, i.e., we first assume that they are satisfied at $n$th order and then prove that they hold for $(n+1)$th order. However, we can note that the condition of Eq. \eq{optimaln} is different for the odd and even orders. Therefore, we need to prove the above results for $n=2k-1$ and $n=2k$ separately.
\\
\\
\noindent
\textbf{(1) Case of $n=2k$}

We assume that the Eqs. \eq{ineqn} and \eq{optimaln} are satisfied when $n=2k-1$ for $k\geq 1$. Taking $(2k-1)$ variation to Eq. \eqref{varid} and considering the null energy condition $\d^{2k} T_{ab}\x^a\x^b\geq 0$ of the perturbation matter fields when the first $(2k-1)$ order perturbation inequalities are saturated, we can further obtain
\ba\begin{aligned}
  \delta^{2k} M-\Omega_H\delta^{2k} J&=\delta^{2k-2}\mathcal{E}_\math{H}(g,\delta g)+\delta^{2k-2}\mathcal{E}_{\Sigma_1}(g,\delta g)+\int_H\tilde{\bm\epsilon}k^a\xi^b\delta^{2k}T_{ab}\\
&\geq\delta^{2k-2}\mathcal{E}_\math{H}(g,\delta g)+\delta^{2k-2}\mathcal{E}_{\Sigma_1}(g,\delta g)\,.
\end{aligned}\ea
For the second term in the right side of the above expression, since $\delta^{2k}\mathcal{E}_{\Sigma_1}(g,\delta g)$ only contains the first $2k$ order variational of $g_{ab}$, it can be evaluated by the one-parameter family $\f_{(2k)}(\a)$ and finally we have
\ba\begin{aligned}
  \delta^{2k-2}\mathcal{E}_{\Sigma_1}(g,\delta g)&=-\frac{\kappa}{8\pi}\delta^{2k} A_{(2k)}\,.
\end{aligned}\ea
From Eq. \eqref{omega}, using the Gaussian null coordinate system \eq{guassiannull}, $\d^{2k-2} \mathcal{E}_\math{H}(g,\delta g)$ can be written as
\ba\begin{aligned}\label{EH2k}
  \delta^{2k-2}\mathcal{E}_\math{H}(g,\delta g)&=-\frac{1}{16 \pi}\delta^{2k-2}\int_\math{H}\bar{\bm\epsilon}\xi_a P^{abcdef}\left(\mathcal{L}_{\xi}\delta g_{bc}\nabla_d\delta g_{ef}-\delta g_{bc}\nabla_d \mathcal{L}_{\xi}\delta g_{ef}\right)\\
  &=-\frac{1}{16 \pi}\delta^{2k-2}\int_\math{H}\bar{\bm\epsilon} \left(\mathcal{L}_{\xi}\delta \gamma_{bc} P^{zbcdef}\nabla_d\delta g_{ef}-\delta \gamma_{bc}P^{zbcdef}\nabla_d \mathcal{L}_{\xi}\delta g_{ef}\right)\,.
\end{aligned}\ea
Using the conditions \eq{optimaln} with $n=2k-1$, especially
\ba\begin{aligned}\label{condition2k}
&\d^j g_{ab}=\d^j\g_{ab}=0\,, \quad\quad\forall 1\leq j\leq k-1\,,\\
&g^{ab}\mathcal{L}_{\xi}\delta^i\gamma_{ab}=\gamma^{ab}\mathcal{L}_{\xi}\delta^i\gamma_{ab}=0\,,\quad\forall\, i\leq 2k-1
\end{aligned}\ea
on $\math{H}$, we can obtain
\ba\begin{aligned}\label{c1}
\d^{m}[g^{bc}\math{L}_{\x}\d \g_{bc}]=\sum_{i=0}^{m}C_{m}^i\d^i(g^{bc})\math{L}_\x \d^{2k-i+1} \g_{bc}=g^{bc}\math{L}_{\x} \d^{m+1}\g_{bc}=0
\end{aligned}\ea
on $\math{H}$ for any $m\leq 2k-2$, where $C_m^i$ is the binomial coefficient. Considering the assumption that the perturbation vanishes at early time, the optimal condition $g^{ab}\math{L}_{\x}\d^i\g_{ab}=0$ implies that
\ba\begin{aligned}
g^{ab}\d^i\g_{ab}=0\,,\quad\quad\forall i\leq 2k-1\,.
\end{aligned}\ea
Then, similar calculation can give $\d^{2k-2}[g^{bc}\d \g_{bc}]=0$. Therefore, we can replace $P^{zbcdef}$ by $\bar{P}^{zbcdef}$. Again using these conditions, from the variation of Eq. \eqref{Pabcdef}, for any integer $0\leq m\leq k-2$, it is not hard to verify
\ba\begin{aligned}\label{1}
\delta^m[\bar{P}^{z}{}_{ij}{}^{def}\nabla_d\delta g_{ef}]&=\delta^m[-\delta\sigma_{ij}+\delta\gamma_{l(i}\Gamma^l{}_{j)u}]\\
&=-\delta^{m+1}\sigma_{ij}+\sum_{i=0}^{m}C_{m}^i\delta^{i+1}\gamma_{l(i}\delta^{m-i}[\Gamma^l{}_{j)u}]=0\,.
\end{aligned}\ea
Similar features are also satisfied for $\delta^m[\bar{P}^{z}{}_{ij}{}^{def}\nabla_d\math{L}_\x\delta g_{ef}]$. Then, \eqref{EH2k} can be simplified to
\ba\begin{aligned}\label{3}
  &\delta^{2k-2}\mathcal{E}_\math{H}=-\frac{C_{2k-2}^{k-1}}{16 \pi}\int_\math{H}\bar{\bm\epsilon} \left[\mathcal{L}_{\xi}\delta^{k} \gamma_{bc} \delta^{k-1}(\bar{P}^{zbcdef}\nabla_d\delta g_{ef})-\delta^{k}\gamma_{bc} \delta^{k-1}(\bar{P}^{zbcdef}\nabla_d\mathcal{L}_{\xi}\delta g_{ef})\right]\,.
\end{aligned}\nn\\\ea
Same calculation as Eq. \eq{1}, we can further obtain
\ba\begin{aligned}
\delta^{k-1}[\bar{P}^{z}{}_{ij}{}^{def}\nabla_d\delta g_{ef}]=-\d^k \s_{ij}\,,\quad\quad \delta^{k-1}[\bar{P}^{z}{}_{ij}{}^{def}\nabla_d \math{L}_{\x}\d g_{ef}]=-\math{L}_{\x}\d^k \s_{ij}\,.
\end{aligned}\ea
Then, straightforward calculation gives
\ba\begin{aligned}
\delta^{2k-2}\mathcal{E}_\math{H}(g,\delta g)
  &=\frac{C_{2k-2}^{k-1}}{16 \pi}\int_\math{H}\bar{\bm\epsilon} \left[\mathcal{L}_{\xi}\delta^k \gamma_{bc} \delta^k\sigma^{bc}-\delta^{k} \gamma_{bc}\mathcal{L}_{\xi}(\delta^k\sigma^{bc})\right]\\
  &=\frac{C_{2k-2}^{k-1}}{8 \pi}\int_{H}\bar{\bm\epsilon} \mathcal{L}_{\xi}\delta^{k} \gamma_{bc} \delta^{k}\sigma^{bc}+\frac{C_{2k-2}^{k-1}}{16\pi}\int_{B_1}\tilde{\bm\epsilon}\delta^{k} \gamma_{bc}\delta^{k}\sigma^{bc}\\
  &=\frac{C_{2k-2}^{k-1}}{8 \pi}\int_{H}\bar{\bm\epsilon} \mathcal{L}_{\xi}\delta^{k} \gamma_{bc} \delta^{k}\sigma^{bc}\geq 0\,,
\end{aligned}\ea
where we used the stability condition $\d^{k}\s_{ab}=0$ on $B_1$.

So far, we have proved $\delta^{2k-2}\mathcal{E}_\math{H}\geq 0$ under the condition \eq{optimaln} with $n=2k-1$. The $(2k)$th-order perturbation inequality becomes
\ba\begin{aligned}
\delta^{2k} M-\Omega_H\delta^{2k} J\geq -\frac{\k}{8\p}\d^{2k} A_{(2k)}\,.
\end{aligned}\ea

Next, we need to show that the saturation of the $(2k)$th order perturbation inequality can give the results in Eq. \eq{optimaln} with $n=2k$. From the above calculation, it is easy to see that saturation of the $(2k)$th inequality needs the following additional conditions
\ba\begin{aligned}
\d^{2k}T_{ab}\x^a\x^b=0\,,\quad\quad \d^k\s_{ab}=0
\end{aligned}\ea
on $\math{H}$. From the Raychaudhuri equation \eq{Ray}, we have
\ba\begin{aligned}
  \frac{d\delta^{2k}\vartheta}{d u}-\k \d^{2k}\vartheta =-\frac{1}{2}\sum_{i=0}^{2k}C_{2k}^i\delta^{i}\vartheta\delta^{2k-i}\vartheta
  -\sum_{i=0}^{2k}C_{2k}^i\delta^i\sigma_{ab}\delta^{2k-i}\sigma^{ab}-\xi^a\xi^b \delta^{2k}T_{ab}=0\,,
\end{aligned}\ea
where we have used the condition \eq{optimaln} with $n=2k-1$ and $\d^{2k}T_{ab}\x^a\x^b=0$ on $\math{H}$. Considering the fact that the perturbation vanishes at early times, this implies $\delta^{2k+2}\vartheta=0$ on $\math{H}$. Together with the condition \eq{optimaln} of $n=2k-1$, we can easily to see that saturation of the first $(2k)$ order perturbation inequalities is equivalent to the condition \eq{optimaln} with $n=2k$. We have completed the proof of the case with $n=2k$.
\\

\noindent
\textbf{(2) Case of $n=2k+1$}

Next, we will prove Eq. \eq{ineqn} and \eq{optimaln} with $n=2k+1$ when the condition \eq{optimaln} with $n=2k$ is satisfied. Same calculation can give

\ba\begin{aligned}
  \delta^{2k+1} M-\Omega_H\delta^{2k+1} J&=\delta^{2k-1}\mathcal{E}_\math{H}(g,\delta g)+\delta^{2k-1}\mathcal{E}_{\Sigma_1}(g,\delta g)+\int_H\tilde{\bm\epsilon}k^a\xi^b\delta^{2k+1}T_{ab}\\
&\geq-\frac{\k}{8\p}\d^{2k+1}A_{(2k+1)}+\delta^{2k-1}\mathcal{E}_\math{H}(g,\delta g)\,,
\end{aligned}\ea
where we have used the null energy condition of the perturbation matter field $\d^{2k+1}T_{ab}\x^a\x^b\geq 0$. For the last term, we have
\ba\begin{aligned}\label{EH2k1}
  \delta^{2k-1}\mathcal{E}_\math{H}=-\frac{1}{16 \pi}\delta^{2k-1}\int_\math{H}\bar{\bm\epsilon} \left(\mathcal{L}_{\xi}\delta \gamma_{bc} P^{zbcdef}\nabla_d\delta g_{ef}-\delta \gamma_{bc}P^{zbcdef}\nabla_d \mathcal{L}_{\xi}\delta g_{ef}\right)\,.
\end{aligned}\ea

Using the condition \eq{optimaln} with $n=2k$, especially
\ba\begin{aligned}\label{condition2k}
&\d^j g_{ab}=\d^j\g_{ab}=0\,, \quad\quad\forall 1\leq j\leq k\,,\\
&g^{ab}\mathcal{L}_{\xi}\delta^i\gamma_{ab}=\gamma^{ab}\mathcal{L}_{\xi}\delta^i\gamma_{ab}=0\,,\quad\forall\, i\leq 2k
\end{aligned}\ea
on $\math{H}$, Using these conditions as well as a same calculation as Eqs. \eq{c1} and \eq{1}, we can further obtain
\ba\begin{aligned}
\d^m[g^{bc}\math{L}_\x \d\g_{bc}]=\d^m[g^{bc}\d\g_{bc}]=0\,,\quad\quad\forall m\leq 2k-1
\end{aligned}\ea
on $\math{H}$, and therefore
\ba\begin{aligned}
\delta^m[P^{z}{}_{ij}{}^{def}\nabla_d\delta g_{ef}]=\delta^m[P^{z}{}_{ij}{}^{def}\nabla_d\math{L}_\x\delta g_{ef}]=0\,,\quad\quad\forall m\leq k-1
\end{aligned}\ea
on $\math{H}$. Using the above results, from Eq. \eq{EH2k1}, it is not hard to see
\ba\begin{aligned}
\delta^{2k-1}\mathcal{E}_\math{H}(g, \d g)=0\,.
\end{aligned}\ea
Finally, the $(2k+1)$th-order perturbation inequality can be written as
\ba\begin{aligned}
  \delta^{2k+1} M-\Omega_H\delta^{2k+1} J\geq-\frac{\k}{8\p}\d^{2k+1}A_{(2k+1)}\,.
\end{aligned}\ea
Saturations this inequality demands the additional condition
\ba
\d^{(2k+1)}T_{ab}\x^a\x^b=0
\ea
on $\math{H}$. Using the condition \eq{optimaln} with $n=2k$ and the Raychaudhuri, straight forward calculation can show that the above condition also gives $\d^{(2k+1)}\vartheta=0$ and finally proved that the saturation of the first $(2k+1)$ order perturbation inequality is equivalent to the condition \eq{optimaln} with $n=2k+1$, i.e., we have completed the proof. $\Box$\\
\\

Finally, we turn to calculate the quantity $\k\d^{n} \bar{A}_{(n)}$, where we define $\bar{A}=A/8\p$. From the expression \eq{AWk} of the area $A$, we have the identity
\ba\begin{aligned}\label{idA}
\bar{A}_{(n)}^2(\a)-2\bar{A}_{(n)}(\a)M_{(n)}^2(\a)-J_{(n)}^2(\a)=0\,.
\end{aligned}\ea
in which these quantity is evaluated in the configuration $\f_{(n)}(\a)$. In this case, we have the setup such that $\d^{i} M_{(n)}=\d^{i} M$ and   $\d^{i}J_{(n)}=\d^{i} J\quad \forall i\leq n-1$, and thus we have
\ba\begin{aligned}
\delta^{i}\bar{A}=\delta^{i}\bar{A}_{(n)}\,,\quad\quad \forall i\leq n-1\,.
\end{aligned}\ea
The saturation condition of the first $(n-1)$ order perturbation inequalities $\delta^i\vartheta=0$ for any $i\leq n-1$ implies that $\delta^i\bar{A}=0\,, \forall i\leq n-1$, i.e., we also have $\d^i \bar{A}_{(n)}=0\,, \forall i\leq n-1$. Using the above results, we can further obtain
\ba\begin{aligned}
\frac{\k}{8\p} \d^{n}A_{(n)}=\frac{1}{2}\sum_{i=1}^{n-1}C_n^i\left(\d^i M\d^{n-i}M-\frac{1}{2(1+\epsilon)}\d^i J \d^{n-2}J\right)\,.
\end{aligned}\ea

\subsubsection{Gedanken experiments to overspun the black hole at higher-order approximation}
\begin{figure}
\centering
\includegraphics[width=0.8\textwidth]{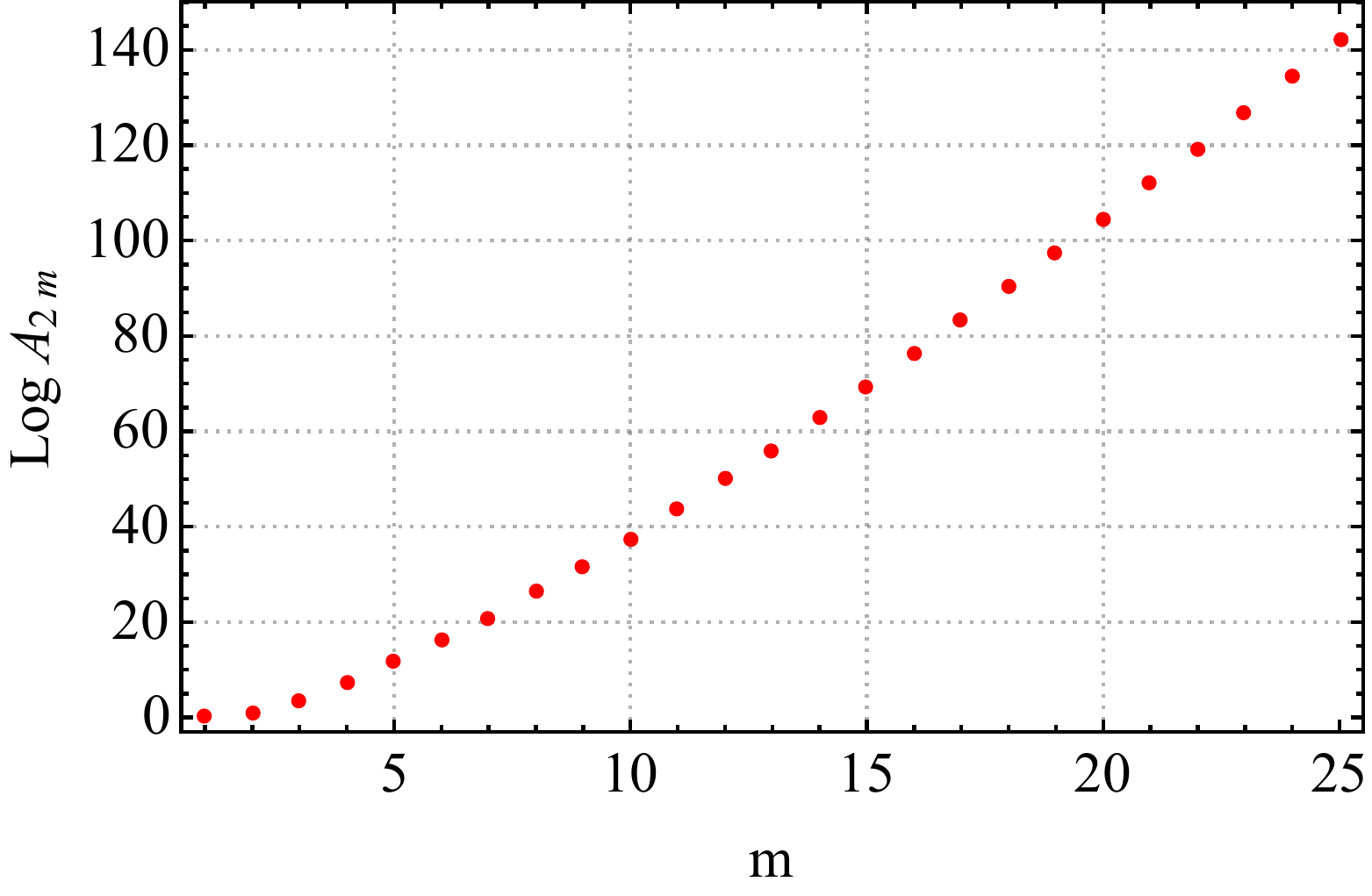}
\caption{Plot showing the value of $A_{2m}$ with different integer $m$}\label{fig2}
\end{figure}

After obtaining the above results, it is not hard to extend the discussion into any arbitrary high-order approximations. With a straightforward extension, we evaluated $h(\l)$ under the first $100$ order approximation of $\lambda$ and $\epsilon$. After considering the perturbation inequality and optimal conditions(i.e., the lower approximation of $h(\l)$ vanishes), we find that the $n$th order approximation of $h(\lambda)$ can always be expressed as
\ba\begin{aligned}
&h(\lambda)\geq 0\,,\quad\forall\, n=2k+1\\
&h(\lambda)\geq B_{n/2}(\lambda^{n/2}\delta^{n/2}J-A_{n/2}\epsilon^{n/2})^2\geq 0\quad\forall\, n=2k\,,
\end{aligned}\ea
with the non-negative integer $k$. Here $B_{n/2}$ is a positive quantity and $A_{2m+1}=0$ for any integer $m\geq 1$. In \fig{fig2}, we show the value of the parameter $A_{2m}$ of different integer $m$. These results show that the nearly extremal Kerr black hole cannot be overspun by the perturbation matter fields with the null energy condition under the higher-order approximation of perturbation. It indicates that the WCCC in Kerr black holes might be straightly valid under the perturbation process.

\section{Conclusion}\label{sec5}
In this paper, we extended the Sorce-Wald gedanken experiments into the higher-order approximations to study the WCCC of a Kerr black hole under the perturbation process. First of all, under the assumptions that the spacetime is stable under the perturbation and the matter fields satisfy the null energy condition, we derived the first four order perturbation inequalities under the corresponding lower optimal conditions. After considering these constraints, we showed that the nearly Kerr black hole cannot be overspun under the fourth-order approximation of perturbation. Then, based on the mathematical induction, we proved that the $n$th-order perturbation inequality can be generally expressed as
\ba\begin{aligned}
\delta^n M-\Omega_H\delta^n J\geq-\frac{\kappa}{8\pi}\delta^n A_{(n)}
\end{aligned}\ea
when the first $(n-1)$th order perturbation inequalities are saturated.
Using the perturbation inequalities, we extended the Sorce-Wald discussion to the first 100 order approximation. Our result showed that the WCCC is always valid in the perturbation process as long as the matter fields satisfy the null energy condition, which implies that the WCCC might be strictly true under the non-approximation level of the perturbation.

\end{document}